# Photoluminescence sidebands of carbon nanotubes below the bright singlet excitonic levels: Coupling between dark excitons and K-point phonons

Yoichi Murakami[1*], Benjamin Lu[2], Said Kazaoui[3], Nobutsugu Minami[4], Tatsuya Okubo[1], Shigeo Maruyama[5*]

[1] *Department of Chemical System Engineering, The University of Tokyo, Bunkyo-ku, Tokyo 113-8656, Japan*

[2] *Department of Bioengineering, Rice University, Houston, Texas 77005, USA*

[3] *Nanotube Research Center, National Institute of Advanced Industrial Science and Technology, Tsukuba, Ibaraki 305-8565, Japan*

[4] *Nanotechnology Research Institute, National Institute of Advanced Industrial Science and Technology, Tsukuba, Ibaraki 305-8565, Japan*

[5] *Department of Mechanical Engineering, The University of Tokyo, Bunkyo-ku, Tokyo 113-8656, Japan*

\* Corresponding authors:

ymurak@chemsys.t.u-tokyo.ac.jp, maruyama@photon.t.u-tokyo.ac.jp

**Abstract**

We performed detailed photoluminescence (PL) spectroscopy studies of three different types of single-walled carbon nanotubes (SWNTs) by using samples that contain essentially only one chiral type of SWNT, (6,5), (7,5), or (10,5). The observed PL spectra unambiguously show the existence of an emission sideband at ~ 145 meV below the lowest singlet excitonic ($E_{11}$) level, whose identity and origin are now under debate. We find that the energy separation between the $E_{11}$ level and the sideband is almost independent of the SWNT diameter. Based on this, we ascribe the origin of the observed sideband to coupling between K-point LO phonons and dipole-forbidden dark excitons, as recently suggested based on the measurement of (6,5) SWNTs.

The optical and electronic properties of one-dimensional (1-D) materials have been an important subject in the field of condensed matter physics [1,2]. The enhanced Coulomb interactions in 1-D often lead to the formation of strongly bound electron-hole (*e-h*) pairs called excitons [3]. It is known that the optical properties of 1-D semiconductors are inherently dominated by excitons [4,5], distinctly different from higher dimensional materials. The sharp spatial localization of strongly bound excitons often makes the interaction between excitons and phonons (*ex-ph* interaction) important [6].

Single-walled carbon nanotubes (SWNTs) [7] are novel 1-D materials that possess various unique properties [8]. The strong confinement in the radial direction (~ 1 nm) and the weak dielectric screening inside SWNTs give rise to very large exciton binding energies of ~ 0.5 eV [9-13], which is much larger than those of InGaAs quantum wires [5,14] and larger than or comparable to those of π-conjugated polymers [15,16]. So far, various groups have studied the *ex-ph* interactions in SWNTs both in terms of theory [17,18] and experiment [19-22], and have revealed the existence of a phonon sideband approximately 200 meV above the energy level of the singlet bright exciton with *s*-envelope (which is a "1u" state based on the symmetry [12,13] but we simply call it "$E_{ii}$" hereafter for the *i*th single-particle subband) [17-22].

At the same time, the techniques of preparing SWNT samples for optical studies have been rapidly improving recently. The technique that utilizes fluorene polymers to disperse SWNTs in organic solvents reported by Nish et al. [23] is considered to be particularly important because of the strong selectivity to specific chiral types of SWNTs. The density gradient separation method reported by Arnold et al. [24,25] is another important technique that separates SWNTs according to their diameters as well as to their metallic/semiconducting property.

In this communication, we performed detailed photoluminescence (PL) spectroscopy studies of three different types of SWNTs ((6,5), (7,5), and (10,5)) using samples containing virtually only one type of SWNT, and unambiguously show the existence of emission sidebands at approximately 145 meV below the $E_{11}$ level for all of these SWNT types. The origin of this sideband is currently under debate. Torrens et al. [26] recently suggested, based on optical measurement of (6,5) SWNTs, that this sideband is due to coupling between dipole-forbidden "dark" excitons with finite angular momentum and K-point LO phonons, which is the same mechanism Perebeinos et al. [17] had predicted for the origin of the phonon sideband approximately 200 meV above the $E_{11}$ level. On the other hand, Kiowski et al. [27] have claimed that these weak emission sidebands arise from "deep-dark excitonic levels", based on their observation that the energy separation between sidebands and corresponding $E_{11}$ levels ("$\delta$" hereafter) was remarkably dependent on the diameter of the SWNT ("$d_{SWNT}$" hereafter). Eventually, the interpretation of our experimental results agrees with the discussion presented by the former.

We prepared three types of samples, denoted as "DG", "PFO", and "PFO-BT". In essence, the first sample was prepared based on the density-gradient separation method [25,28], and the latter two samples utilized fluorene polymers to individually disperse SWNTs in toluene [23]. The sample preparation procedures are described as follows.

The preparation of the DG sample started by dispersing ACCVD SWNTs [29,30] at 2 mg/mL in $D_2O$ with 0.33 wt% sodium deoxycholate surfactant. After horn sonication

(Hielscher, UP400s) at 80% duty ratio for 30 min, the mixture was centrifuged at 275,000g for 1 h. The upper 50% was collected and combined with a 1% sodium dodecyl sulfate (SDS), 1% sodium cholate (SC), and 40% iodixanol (IDX) solution. Gradient layers consisting of IDX (20-40%), 1.5% SDS, and 1.5% SC were manually laid, placing the SWNT solution on top of the 40% IDX layer. The samples were then centrifuged at 197,000g for 20 h. The violet-colored fraction in the centrifuge tube was extracted using a micropipette.

The PFO sample was started from a mixture of PFO polymer (American Dye Source Inc., ADS129BE), toluene, acetic acid, and CoMoCAT SWNTs (SouthWest NanoTech. Inc., batch 11L), whose amounts were 6 mg, 7 g, 1 g, and 10 mg, respectively. The addition of acetic acid is to suppress the amounts of (6,5) and (8,3) tubes relative to (7,5) tubes in the obtained sample [31, 32]. The mixture was sonicated in a water bath sonicator for 3 h and then more strongly sonicated with a horn-type ultrasonicator for 15 min at 50% duty ratio. The sonicated liquid was immediately centrifuged at 20,000g for 5 min and the upper 50% volume was collected for the measurement.

The "PFO-BT" sample was prepared using PFO-BT polymer (American Dye Source Inc., ADS133YE), SWNTs (HiPco, Carbon Nanotechnologies, Inc.) and toluene, in the following ratio: 5 mg, 5 mg, and 30 ml, respectively. The mixture was first sonicated with a water bath sonicator for 1 h, followed by a horn-type ultrasonicator for 5 min, and then centrifuged at 150,000g for 60 min. Finally, the upper 80% of the supernatant solution was collected for further characterization.

Figure 1 shows colored contour maps of the PL excitation intensities measured from the (a, d) DG, (b, e) PFO, and (c, f) PFO-BT samples. Panels (d) - (f) are the same as panels (a) - (c), but the intensities are presented in log-scale. This figure shows that the samples contained virtually only one chiral type of SWNTs while the amounts of other types of SWNTs were highly suppressed. Panels (d) - (f) clearly show the existence of sidebands (indicated by the arrows) associated with the PL features of (6,5), (7,5), and (10,5) SWNTs, respectively. These sidebands look very similar to those reported by Kiowski et al. [27], however, there is a qualitative difference in the observed character as described later.

Figure 2 shows PL spectra measured from the (a) DG, (b) PFO, and (c) PFO-BT samples, obtained by resonantly exciting the $E_{22}$ levels of the dominant SWNTs with excitation wavelengths 570, 655, and 800 nm, respectively. The intensities are shown in log scale and the abscissa shows the energy relative to the $E_{11}$ peak energy of the dominant SWNTs in the sample. The peaks indicated by asterisks correspond to the sidebands shown in Fig. 1. Spectra (a) and (b) in Fig. 2 show additional features at approximately 50 meV below the $E_{11}$ levels, as indicated by crosses [33].

Figure 3(a) shows the dependence of the peak widths (FWHM) of the observed $E_{11}$ PL emission features (solid circles) and sideband features (solid squares) on $d_{SWNT}$. The decrease of the widths of the $E_{11}$ PL emission features with the increase of $d_{SWNT}$ agrees with the behavior reported by Inoue et al. [34]. Although the samples measured in this study were ensembles of SWNTs, the clear dependence of the widths of the $E_{11}$ PL emission features on $d_{SWNT}$ (solid circles in Fig. 3(a)) indicates that these were not completely smeared out by the ensemble effect. In fact, the widths of the $E_{11}$ PL peaks (15 - 30 meV) are approximately twice those measured from a single, isolated SWNT at room temperature [34]. It is noted that the widths of the sideband features are always larger than those of the $E_{11}$ features, which indicates that these sidebands are not artifacts due to $E_{11}$ PL emissions from other SWNTs with larger diameters.

The filled squares plotted in Fig. 3(b) represent the dependence of $\delta$ on $d_{SWNT}$, which shows that the observed $\delta$ is virtually independent of $d_{SWNT}$ within the ± 5 meV measurement uncertainty. As a comparison, the values of $\delta$ reported for the similar sidebands in Ref. [27] are plotted with open triangles in the same figure and show a remarkable dependence on $d_{SWNT}$, contrary to our result. Kiowski et al. [27] explained that the observed PL sidebands were weak PL emission from "deep-dark excitonic states" below the $E_{11}$ levels, even lower than the energy level of the singlet dark exciton with zero angular momentum and $s$-envelope (which is termed the "1g" exciton [12,13] whose energy is ~ 5 meV lower than the $E_{11}$ (or "1u") level [35-37]). It is noted that these authors have ruled out the possibility of phonon-involved processes regarding the origin of these sidebands based on their observation that $\delta$ was remarkably dependent on $d_{SWNT}$ [27].

In contrast, the independence of $\delta$ on $d_{SWNT}$ observed in this study suggests an involvement of phonons in the origin of these sidebands, particularly the involvement of higher-frequency (> 1000 cm$^{-1}$) phonons that arise from the phonon dispersion relations of graphite [38], and hence is essentially independent of $d_{SWNT}$. So far, the phonon sideband at ~ 200 meV above the $E_{ii}$ levels has been well recognized and studied [17-22]. This phonon sideband was first predicted by Perebeinos et al. [17] as a result of strong coupling between dipole-forbidden excitons and K-point LO phonons (~ 170 meV). It is noted that the same LO phonon is responsible for the D-band in Raman scattering from graphite and carbon nanotubes [17,19].

Figure 4 shows a schematic diagram that describes the four lowest-energy states of singlet excitons in SWNTs [35]. The abscissa ($q$) denotes the momentum of the excitons. The solid (dashed) curve represents the dipole-allowed (dipole-forbidden) excitonic state with zero angular momentum $E_{11}$ ($E_{1g}$). The dot-dashed curve denotes the dipole-forbidden exciton band with finite angular momentum, whose energy is approximately 20 – 30 meV higher than the $E_{11}$ level at $q = 0$ [35]. Only the excitons at $q \approx 0$ in the $E_{11}$ level can directly interact with photons. We have shown here the diagram for zigzag SWNTs [35,39] for simplicity, but the generality in the following discussion is not affected by this simplification. Based on this schematic, the phonon sideband predicted by Perebeinos et al. corresponds to the point 'A', emerging as a result of the scattering of the dark exciton at $q = q_0$ (point 'X') by K-point LO phonons with $q = - q_0$ [17].

We explain the origin of the observed sideband at ~ 145 meV below $E_{11}$ as follows, which eventually coincides with the interpretation suggested by Torrens et al. [26]. The strong coupling between dark excitons with finite angular momentum and K-point LO phonons (~ 170 meV $\equiv E_{ph}$) predicted by Perebeinos et al. [17] also causes the scattering of dark excitons at $q = q_0$ (at 'X' in Fig. 4) into the bright excitonic state at $q = 0$ ('B' in Fig. 4), reducing their energy by $E_{ph}$ through an emission of a K-point LO phonon of wavevector $q_0$, *satisfying both the energy and momentum conservation requirements simultaneously*. In this process, the phonon sideband is expected to appear at approximately {170 meV - 25 meV} = 145 meV below the $E_{11}$ level, which agrees well with our experimental results. We note that the sidebands disappeared at 6 K according to the results shown by Kiowski et al. [27], and this seems to support the above phonon-mediated interpretation, although the qualitative discrepancy shown in Fig. 3(b) still remains a question.

In summary, we have performed detailed PL studies of (6,5), (7,5), and (10,5) SWNTs by using samples that contain essentially only one chiral type of SWNT. Their PL spectra unambiguously show sideband features at ~ 145 meV below $E_{11}$ levels for all of these SWNT types. The experimental results showed that $\delta$ was almost independent of $d_{\text{SWNT}}$ within the resolution of our data (± 5 meV). This result suggests that the observed sidebands emerge as a result of strong coupling between dark excitons with finite angular momentum and K-point LO phonons, in agreement with the recent suggestion by Torrens et al. [26] based on the measurement of (6,5) SWNTs.

**Acknowledgement**

Part of this work was financially supported by Grant-in-Aid for Scientific Research (19206024 and 19054003) from the Japan Society for the Promotion of Science, SCOPE (051403009) from the Ministry of Internal Affairs and Communications, NEDO (Japan), and MITI's Innovation Research Project on Nanoelectronics Materials and Structures. One of the authors (Y. M.) was financially supported by the JSPS grant #18-09883. B. L. worked as a participant in the Rice University NanoJapan 2008 program, sponsored by a Partnership for International Research & Education Grant through the U.S. National Science Foundation.

**Figure Captions**

FIG. 1: (Color online) Contour maps of the PL excitation intensities measured from the (a, d) DG, (b, e) PFO, and (c, f) PFO-BT samples. Intensity scale-bars are attached to the right of each panel, where the unit is photon count. Panels (d) - (f) are the same as panels (a) - (c), but the intensities are shown in log scale. Arrows indicate the locations of the sidebands.

FIG. 2: (Color online) PL spectra measured from the (a) DG, (b) PFO, and (c) PFO-BT samples, obtained by resonantly exciting the $E_{22}$ levels of the dominant SWNTs with excitation wavelengths of 570, 655, and 800 nm, respectively. The abscissa represents the energy relative to the $E_{11}$ levels of (6,5), (7,5), and (10,5) SWNTs. Asterisks (*) and crosses (+) indicate the observed sidebands and shoulders of $E_{11}$ PL emission features, respectively.

FIG. 3: (Color online) (a) Dependence of the peak widths (FWHM) of the $E_{11}$ PL emissions (solid circles) and the observed sidebands (solid squares) on $d_{SWNT}$. For the latter, the error bars account for a ± 3 meV measurement uncertainty. (b) Solid squares: The measured relationship between $\delta$ and $d_{SWNT}$ in this study. Error bars account for a ± 5 meV uncertainty. Open triangles: The relationship between $\delta$ and $d_{SWNT}$ reported for similar sidebands in Ref. [27].

FIG. 4: (Color online) Schematic diagram showing the four lowest-energy states of singlet excitons in SWNTs. The abscissa denotes the momentum of the excitons.

The solid and dashed curves denote the bright and dark exciton bands with zero angular momentum, termed $E_{11}$ and $E_{1g}$ levels in this report, respectively. The dot-dashed curve denotes the dark exciton band with non-zero angular momentum. The arrow connecting the points 'X' and 'B' shows the process suggested for the PL sidebands observed in this study.

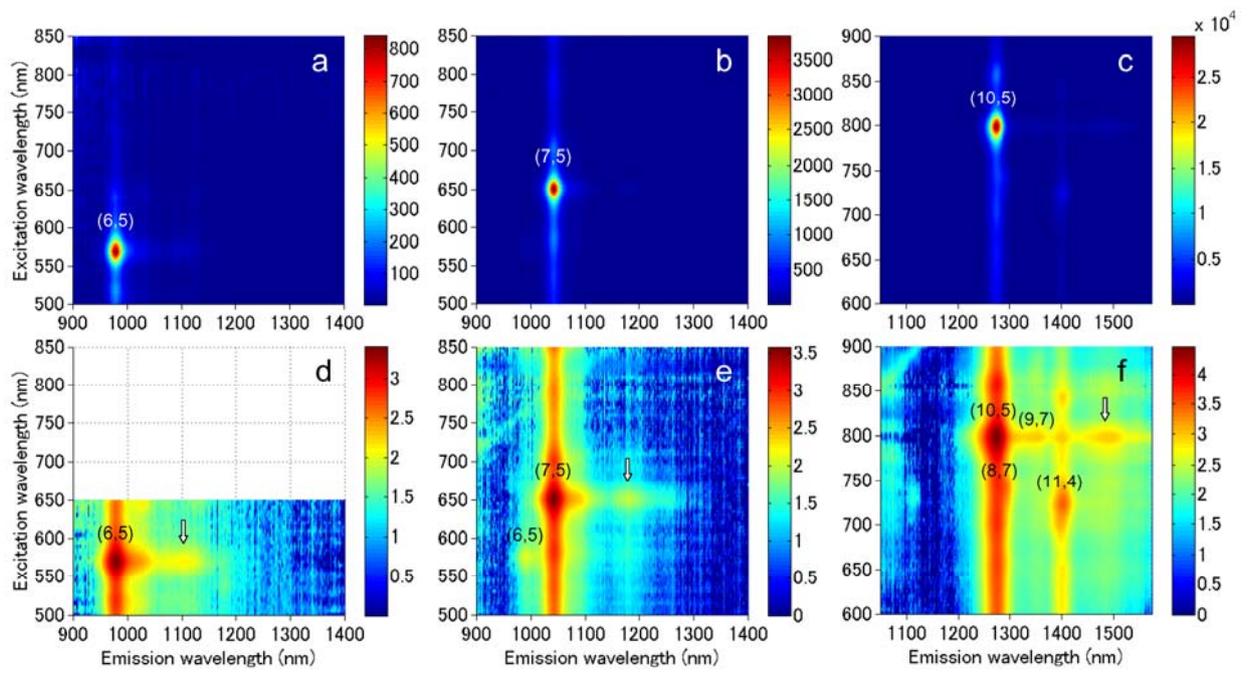

Figure 1

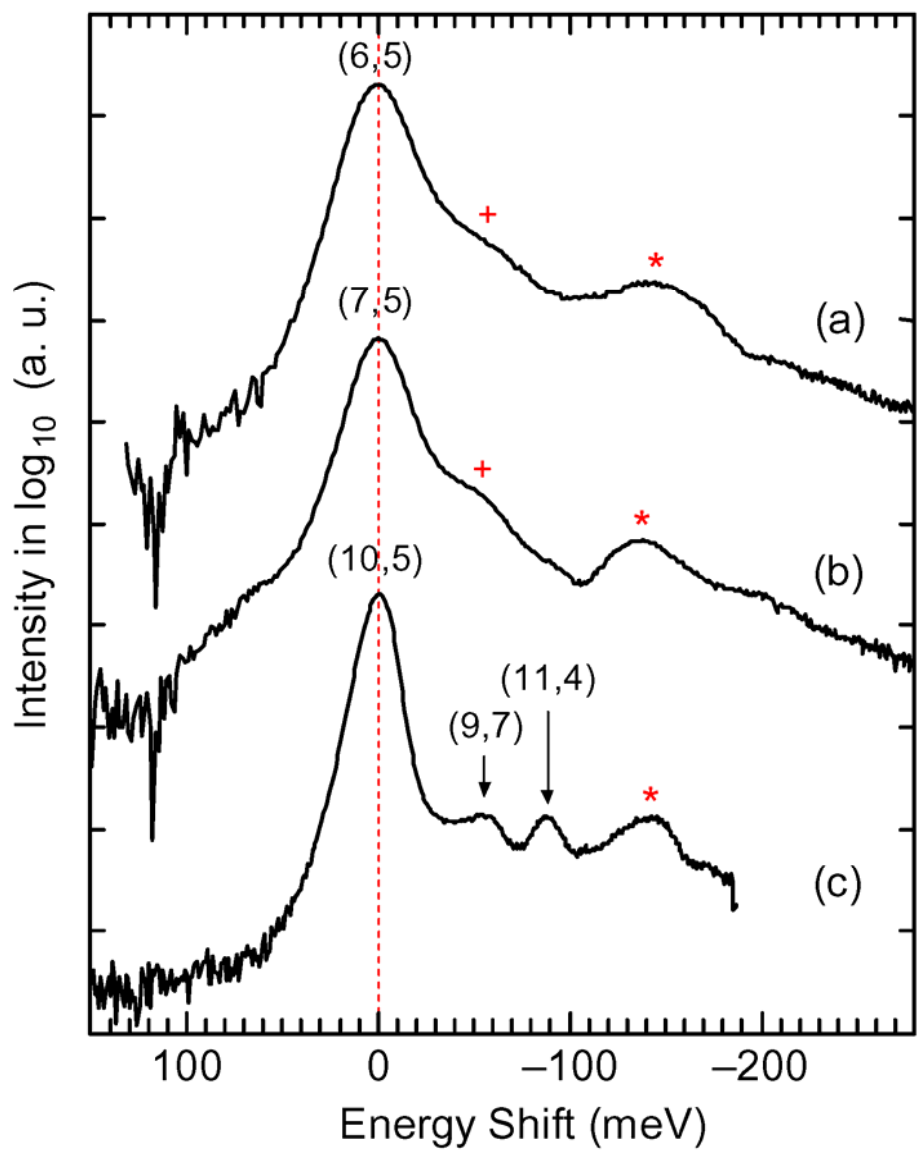

Figure 2

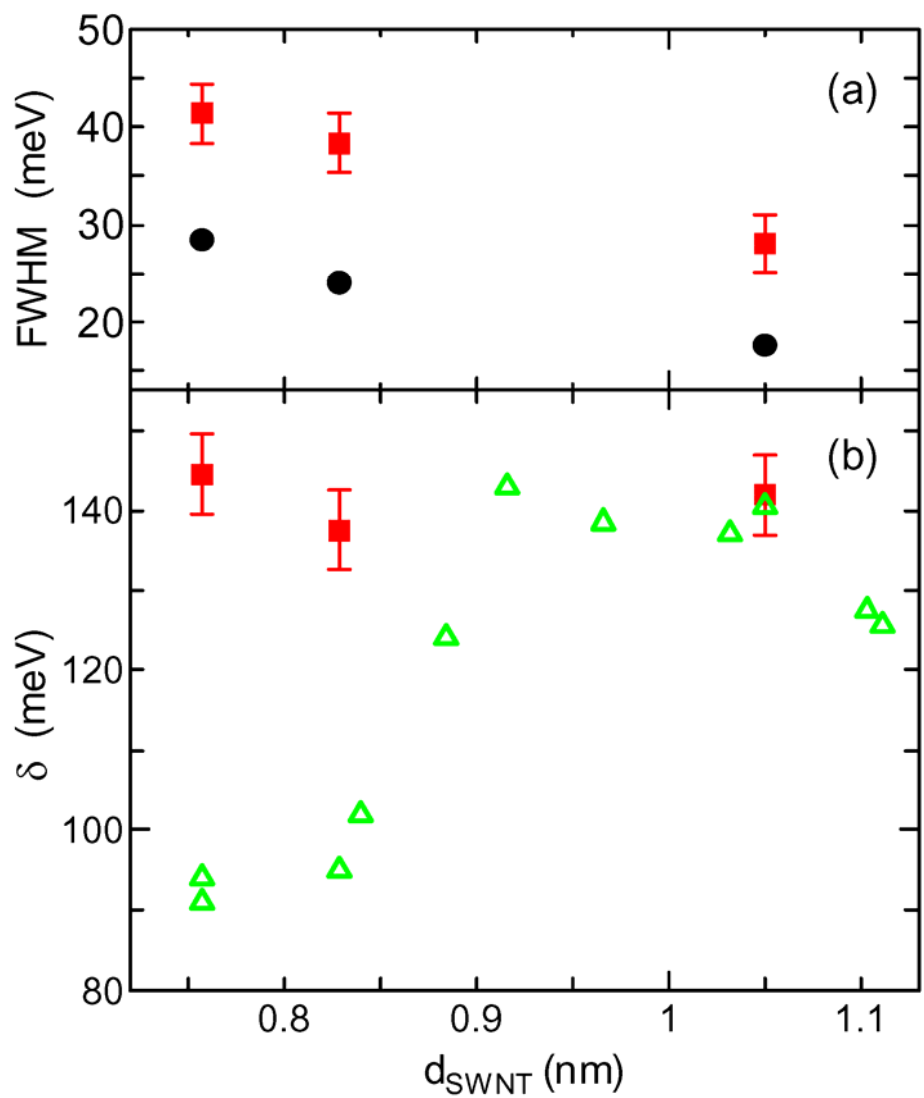

Figure 3

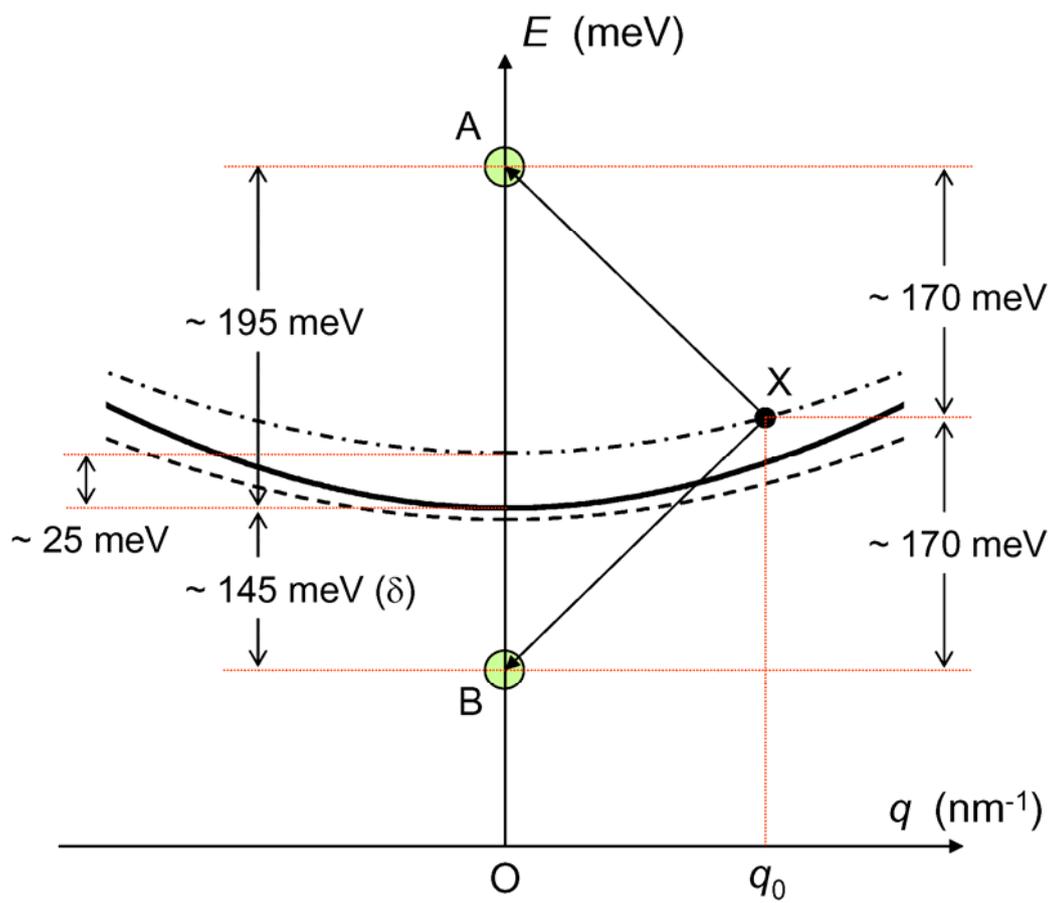

Figure 4

# Supplementary Information

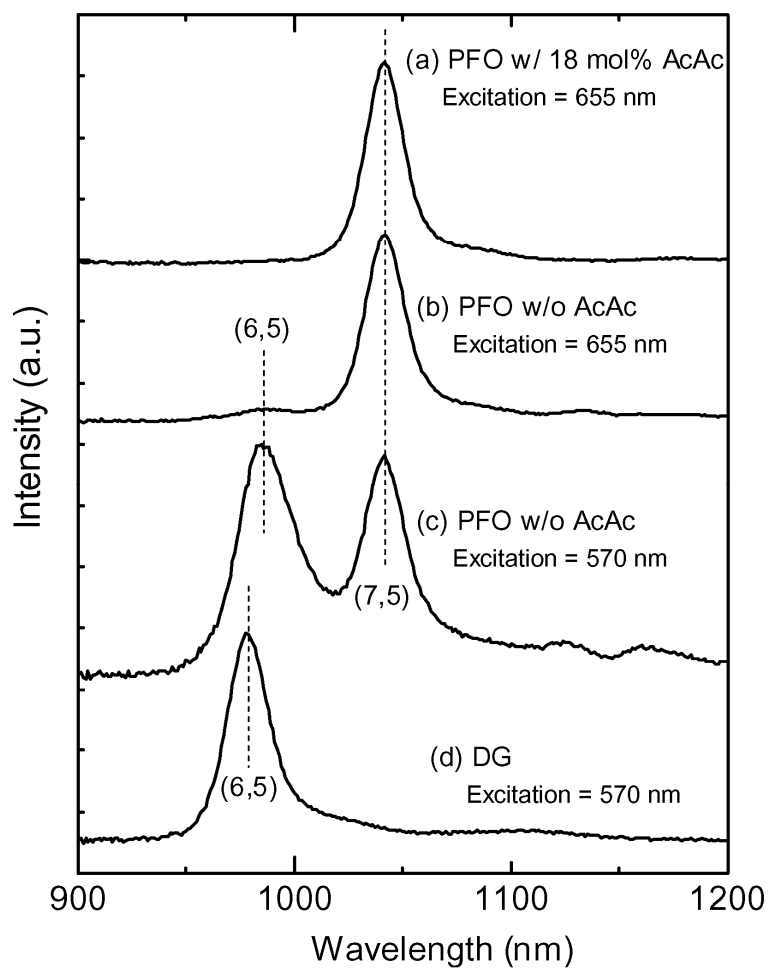

**Figure S1.**

This figure compares the PL emission spectra of (a) the PFO sample that was studied in this report (shown in Figs. 1 and 2) that contains 18 mol% acetic acid measured with 655 nm excitation light, (b) the PFO sample prepared without the addition of acetic acid measured with 655 nm excitation light, (c) the same sample as (b) but measured with 570 nm excitation light, and (d) the DG sample measured with 570 nm excitation light. This figure indicates that the $E_{11}$ energy of (7,5) tubes is not changed by the addition of acetic acid. There is a small energy difference (~ 9 meV) in the $E_{11}$ energies of (6,5) SWNTs between spectra (c) and (d) that is thought to be caused by the difference in their surrounding environments.

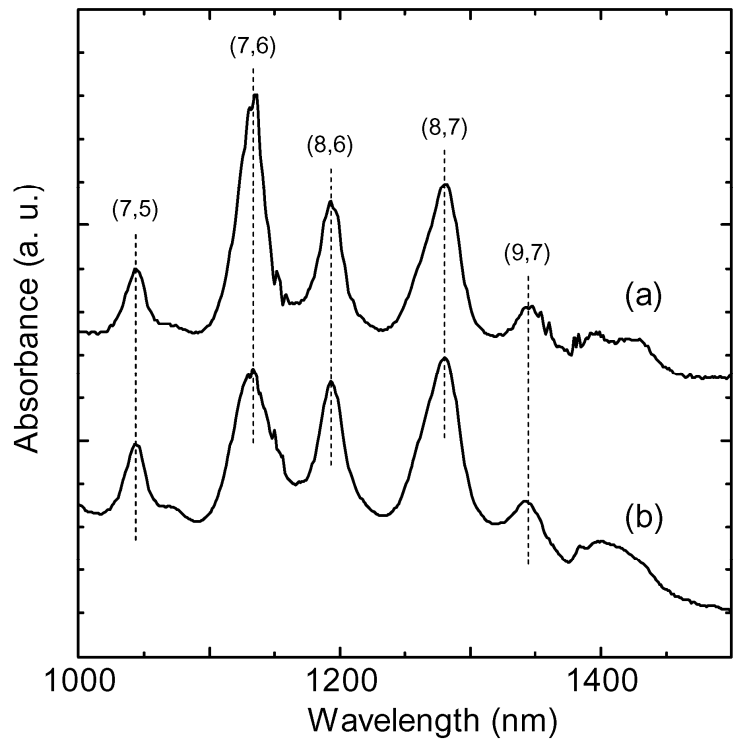

**Figure S2.**

This figure compares the optical absorption spectra measured from (a) the PFO-HiPco sample prepared without addition of acetic acid and (b) the PFO-HiPco sample prepared with 13 mol% of acetic acid. The optical absorption of toluene has been subtracted from the spectra. This figure shows that the addition of acetic acid does not cause the energy shifts of the $E_{11}$ levels also for larger-diameter HiPco SWNTs. It is also noted that the addition of acetic acid does not cause the doping of the SWNTs.